\input harvmac
\def\half{{1 \over 2}}

\def\>{{\rangle}}
\def\<{{\langle}}

\def\p{{\partial}}

\def\s{{\sigma}}

\def\l{{\lambda}}

\def\a {{\alpha}}
\def\b {{\beta}}

\def\ad {{\dot \a}}
\def\bd {{\dot \b}}

\def\d {{\delta}}

\def\e {{\epsilon}}

\def\t {{\theta}}

\def \t {{\theta}}

\def \tb {{\bar\theta}}

\Title{\vbox{\hbox{IFT-P.055/97}}}
{\vbox{\centerline{{\bf Construction of $R^4$ Terms in N=2
D=8 Superspace}}}}  
\bigskip\centerline{Nathan Berkovits}
\bigskip\centerline{Instituto
de F\'{\i}sica Te\'orica, Univ. Estadual Paulista}
\centerline{Rua Pamplona 145, S\~ao Paulo, SP 01405-900, BRASIL}
\bigskip\centerline{e-mail: nberkovi@ift.unesp.br}
\vskip .2in
Using linearized superfields, $R^4$ terms in the Type II superstring 
effective action compactified on $T^2$ are constructed as integrals 
in N=2 D=8 superspace. The structure of these superspace integrals 
allows a simple proof of the $R^4$ non-renormalization theorems which 
were first conjectured by Green and Gutperle.      

\Date{September 1997}
\newsec {Introduction}

One way to verify that M-theory provides a consistent non-perturbative
definition of superstring theory is to compare M-theory predictions with
superstring scattering amplitudes. In general, this comparison is difficult
since one needs to know the non-perturbative behavior of the superstring
amplitudes. 

However, one special case where comparison is possible is the $R^4$ term
in the Type II superstring effective action. For the Type IIB superstring,
this term only receives tree-level, one-loop, and non-perturbative corrections,
while for the Type IIA superstring, it only recieves tree-level and
one-loop contributions. This loop-dependence of $R^4$ terms was first
conjectured on the basis of SL(2,Z) symmetry\ref\Gr{M. Green and
M. Gutperle, Nucl. Phys. B498 
(1997) 195, hep-th/9701093.}
and was later verified by
comparison with 
$R^2$ terms in the 
effective action obtained by compactification on $K3\times T^2$\ref\piol
{I. Antoniadis, B. Pioline and T.R. Taylor, ``Calculable $e^{1/\l}$ Effects'',
hep-th/9707222.}.

The loop-dependence of $R^4$ terms is reminiscent of terms in the 
effective action of the Type II superstring compactified to four dimensions,
where low-energy decoupling of N=2 D=4 vector and hypermultiplets implies 
various
non-renormalization theorems.\ref\str{B. de Wit, P. Lauwers
and A. Van Proeyen, Nucl. Phys. B255 (1985) 269\semi
A. Strominger,
Nucl. Phys. B451 (1995) p.96, hep-th/9504090.}
The easiest way to prove this decoupling is
to note that low-energy terms come from N=2 D=4 superspace actions which
only integrate over half of the eight 
$\theta$'s. Since vector multiplets are represented
by chiral superfields and hypermultiplets (or more accurately, tensor
hypermultiplets) are represented by linear superfields, there is no local
superspace action containing both types of superfields which only 
depends on half
the $\t$'s\ref\eff{N. Berkovits and W. Siegel,
Nucl. Phys. B462 (1996) 213, hep-th/9510106.}. 

As will be shown in this paper, a similar argument can be used to prove
non-renormalization theorems for $R^4$ terms.  This is done by
first constructing N=2 D=8 superspace integrals for
$R^4$ terms in the effective action of the
Type II superstring compactified on $T^2$.  
Although there is only
one irreducible N=2 D=8 multiplet\ref\salam{A. Salam and E. Sezgin,
Nucl. Phys. B258 (1985) 284.}, 
it can be represented at linearized
level as either 
a chiral superfield (whose lowest component 
is a complex scalar parameterizing
SL(2)/SO(2)) 
or as a linear superfield
(whose lowest components are five scalars parameterizing SL(3)/SO(3)).

As in the N=2 D=4 case, the chiral and linear
superfield can not both appear in 
N=2 D=8 superspace actions which integrate over half of the 
thirty-two $\t$'s. This implies that the zero modes of the SL(2)/SO(2)
scalars and SL(3)/SO(3) scalars decouple. 
After constructing $R^4$ terms as N=2 D=8
superspace integrals over sixteen $\t$'s,
this decoupling is used to prove that $R^4$ terms in the effective action
of the uncompactified
Type IIB
superstring only appear at tree-level, one-loop and non-perturbatively, while
$R^4$ terms in the effective action of the uncompactified
Type IIA
superstring only appear at tree-level and one-loop.

\vskip 20pt 
In the second section of this paper,
the proof of non-renormalization theorems for
N=2 D=4 systems is reviewed. In the third section, a chiral and linear
superfield is constructed for the N=2 D=8 supergravity
multiplet at linearized level and N=2 D=8 
superspace actions are constructed for terms with 
eight derivatives.  
In the fourth section, this construction is used to prove 
non-renormalization theorems for $R^4$ terms in the
effective action 
of the uncompactified Type IIA and Type IIB superstrings. In the
fifth section, possible generalizations of the $R^4$ non-renormalization
theorems are discussed. 

\newsec{Review of the N=2 D=4 Action for the Compactified Type II Superstring}

Since the superspace structure of the N=2 D=8 effective action closely
resembles that of the N=2 D=4 effective action, 
the N=2 D=4 superspace effective action\eff for the compactified Type II
superstring will be reviewed first. 

\subsec{ Scalar versus Tensor Hypermultiplets}

The matter superfields present in the N=2 D=4 action consist of vector
multiplets and hypermultiplets. In four dimensions, tensors are on-shell
equivalent to scalars, so one can either represent the hypermultiplets
by tensor hypermultiplets (containing one tensor and three scalars) or
by scalar hypermultiplets (containing four scalars). 
Although most of the literature uses scalar hypermultiplets, N=2 D=4
superspace effective actions are much easier to construct if one
uses tensor hypermultiplets. 

Furthermore, superstring field theory and 
sigma model arguments 
suggest that the correct off-shell
representation is the tensor hypermultiplet rather than the scalar
hypermultiplet. As will be discussed in the following subsection, N=2 D=4 
vector and tensor multiplets are described by chiral/chiral and 
chiral/anti-chiral spacetime superfields. This structure is predicted
by string field theory\ref\sieg{W. Siegel, Phys. Rev. D53 (1996)
3324, hep-th/9510150.}
since a D=4 Type II closed superstring field should 
be the product
of two D=4 open superstring fields, and the massless sector
of the open D=4
superstring includes N=1 D=4 Wess-Zumino scalar multiplets which are described
by chiral and anti-chiral superfields. 

This same superspace structure is also
predicted by the spacetime-supersymmetric sigma model for the D=4 Type II
superstring, which relates superspace chirality
and worldsheet chirality\ref\Be{N. Berkovits, Phys. Lett. B304
(1993) 249, hep-th/9303025.}\eff.
For the Type IIB (Type IIA) superstring, the chiral/chiral spacetime
superfields
for vector multiplets come from chiral/chiral (chiral/anti-chiral)
worldsheet moduli of the Calabi-Yau manifold and the chiral/anti-chiral
spacetime superfields for the tensor hypermultiplets come from
chiral/anti-chiral (chiral/chiral) worldsheet moduli of the Calabi-Yau
manifold. 

At least at the perturbative level, representing some Ramond-Ramond
scalars by tensors does not restrict the action because the R-R zero
modes decouple, so one can obtain the scalar action by performing a
duality transformation on the tensor action.  
At the non-perturbative level,
there could be difficulties when the abelian gauge-invariance of the
tensor is broken to a discrete subgroup in the effective action.\ref\dew
{B. de Wit, private communication.}
However, as discussed in \ref\der{
C.P. Burgess, J.-P. Derendinger, F. Quevedo and
M. Quiros, Phys. Lett. B348 (1995) 428.}, 
it appears possible to describe even these
actions with tensor multiplets. 

\subsec{N=2 D=4 Superfields}

The variables of N=2 D=4 superspace are [$x^\mu$, $\t^\a_j$, $\tb^{j \ad}$]
where $\mu=0$ to 3, $\a$ and $\ad$ = 1 to 2, and $j=1$ to 2 is
an internal $SU(2)_R$ index which is raised and lowered using the
anti-symmetric $\e^{jk}$ tensor. $\tb^{j \ad }$ is the complex conjugate
of $\t^\a_j$, and under $U(1)_R$ transformations,
$\t^\a_j$ carries $+1$ charge and $\tb^{j\ad}$ carries $-1$ charge.

Under supersymmetry transformations parameterized by $\xi^\a_j$ and
$\bar\xi^\ad_j$, 
$$\d\t^\a_j =\xi^\a_j,\quad 
\d\tb^{j\ad} =\bar\xi^{j\ad},\quad 
\d x^\mu = i\s^\mu_{\a\ad}(\xi^\a_j\tb^{j\ad }  
+\bar\xi^{j\ad} \t_j^{\a }) ,$$ 
and supersymmetric derivatives are defined as 
\eqn\aaa{D_\a^j 
= {\p \over {\p \t^\a_j}} + i \tb^{j\ad } \s^\mu_{\a\ad} \p_\mu,\quad
\bar D^\ad_j = 
{\p \over {\p \tb_\ad^j}} + i \t_{j\a} \bar\s_\mu^{\ad\a} \p^\mu.}

The field-strength of a vector multiplet is described by a restricted
chiral 
superfield $W$ satisfying
\eqn\chir{\bar D_\ad^+ W =\bar D_\ad^-W =  0,}
$$ D_\a^{(j} D^{k) \a} W = 
\bar D_\ad^{(j} D^{k) \ad} \bar W $$
where the first constraint implies that $W$ is chiral/chiral, 
while the second constraint
implies that $W$ is restricted. 
The physical bosonic components of $W$ appear as
\eqn\physw{
W = w(x) + \t_j^{(\a} \t^{\b) j} \s^{\mu\nu}_{\a\b} F_{\mu\nu} (x) + ... }
where $w$ is a complex scalar and $F_{\mu\nu}$ is the vector field strength. 
Under $U(1)_R\times SU(2)_R$, $W$ transforms as $(+2,1)$, so 
$w$ and $\bar w$ transform as $(+2,1)$ and $(-2,1)$ while $F_{\mu\nu}$
transforms as $(0,1)$. 

The field-strength of a tensor hypermultiplet is described by
a linear superfield $L_{jk}$ symmetric in its SU(2) indices which
satisfies the reality condition $L_{jk}=(L^{jk})^*$ and the linear constraint 
\eqn\aab{ D^\a_{(j} L_{kl)} = 0, \quad
\bar D^\ad_{(j} L_{kl)} = 0.}
The physical bosonic components of $L_{jk}$ appear as 
\eqn\physl{L_{jk} = l_{jk}(x) + \t_{(j}^\a \tb_{k)}^\ad 
\e_{\mu\nu\rho\kappa}\s^\mu_{\a\ad} H^{\nu\rho\kappa}(x) + 
... }
where $l_{jk}$ is a triplet of scalars transforming as
$(0,3)$ under $U(1)_R\times SU(2)_R$ and $H^{\mu\nu\rho}$ is the tensor
field-strength which transforms as $(0,1)$. 

Although the constraints of \aab appear very different from the
constraints of \chir, they are actually closely related. This can be seen
by noting that the constraints of \aab imply that $L_{++}$ is restricted
twisted-chiral since it satisfies 
\eqn\twist{D_+^\a L_{++} =\bar D^{-\ad} L_{++}=0,}
$$D_-^\a D_{-\a} L_{++} = D_+^\a D_{+\a}
L_{--},\quad 
\bar D_-^\ad \bar D_{-\ad} L_{++} = \bar D_+^\ad \bar D_{+\ad} L_{--}.$$ 
The first two constraints imply that $L_{++}$ is chiral/anti-chiral, while
the second two constraints imply that $L_{++}$ is restricted. 

\subsec{N=2 D=4 Superspace Actions}

Two-derivative actions for $M$ vector multiplets and $N$ tensor
hypermultiplets can be written in manifestly
supersymmetric notation as
\eqn\act{
\int d^4 x|_{\t_j^\a=\tb_j^\ad=0} [(D_+)^2 (D_-)^2 f_V(W^{(I)}) ~+ ~
\oint_0 d\zeta 
(D_-)^2 (\bar D^+)^2 f_T(\tilde L^{(J)},\zeta) ~+~c.c.]}
where $I=1$ to $M$, $J=1$ to $N$, $f_V$ and $f_T$ are arbitrary functions 
of $M$ and $N+1$ variables, 
$\oint_0 d\zeta$ is a contour integration around
$\zeta=0$, and
\eqn\ttt{\tilde 
L^{(J)} =  L^{(J)}_{++} +\zeta L^{(J)}_{+-} + \zeta^2 L^{(J)}_{--}.}
The hypermultiplet 
contribution
to \act is supersymmetric where 
\eqn\uuu{\d_Q f_T = [\xi_j^\a D_\a^j +
\bar\xi^j_\ad \bar D^\ad_j -2i(\xi\s^\mu\tb 
+\bar\xi\bar\s^\mu\t)\p_\mu] f_T,}
since
$D_+^\a f_T = \zeta D_-^\a f_T$ and 
$\bar D^-_\ad f_T = \zeta \bar D^+_\ad f_T$, so 
$(D_-)^2 (\bar D^+)^2 \d_Q f_T$ is a total derivative in $x^\mu$ 
\ref\hyper{A. Karlhede, U. Lindstrom and M. Ro\v cek, Phys. Lett. B147
(1984) 297\semi W. Siegel, Phys. Lett. B153 (1985) 51.}.

Note that integrating over all eight $\t$'s (i.e. using 
$(D_+)^2 (D_-)^2 
(\bar D^+)^2 (\bar D^-)^2 $) would imply a minimum of four derivatives
in the action, assuming the action is local. In this paper, non-local
actions (such as those coming from holomorphic anomalies\ref\hol
{M. Bershadsky, S. Cecotti, H. Ooguri and C. Vafa, Comm.
Math. Phys. 165 (1994) 311, hep-th/9309140\semi 
I. Antoniadis, E. Gava, K.S. Narain and
T.R. Taylor, Nucl. Phys. B413 (1994) 162.}\ref\dix
{L. Dixon, V.S. Kaplunovsky and J. Louis, Nucl. Phys. B355 (1991)
649.} which involve
$(\p^\mu \p_\mu)^{-1}$) will not be discussed. 

As shown in \eff, the above 
action can be easily coupled to supergravity by introducing 
a compensating vector multiplet, a compensating tensor
hypermultiplet, and a physical tensor
hypermultiplet which is the `universal
hypermultiplet'. 

\subsec{N=2 D=4 Non-Renormalization Theorems}

To prove non-renormalization theorems, one uses the fact that 
the zero modes
of Ramond-Ramond fields decouple at the perturbative level.
Ramond-Ramond zero modes only
appear in the lowest components
of $L_{+-}^{(J)}$, so the perturbative contribution to
\act needs to be invariant
under $\tilde L^{(J)} \to 
\tilde L^{(J)} + \zeta C^{(J)}$ where $C^{(J)}$ are constants. This is
true whenever $f_T(\tilde L^{(J)},\zeta)= \zeta^{-1} g(\tilde L^{(J)})$ 
where $g(\tilde L^{(J)})$
is a function of $N$ variables since, in
this case, the shift in $\tilde L^{(J)}$ cancels the pole when $\zeta=0$. 

So at the perturbative level, the hypermultiplet contribution
to the action of \act 
simplifies to 
\eqn\simpli{\int d^4 x|_{\t_j^\a=\tb_j^\ad=0}  
\oint_0 d\zeta \zeta^{-1} (D_-)^2 (\bar D^+)^2 
g(\tilde L^{(J)})~+~c.c.}
$$= \int d^4 x|_{\t_j^\a=\tb_j^\ad=0} 
(D_-)^2 (\bar D^+)^2 
g(L_{++}^{(J)})~+~c.c.~~.$$
It is interesting to
note that if $f_V(W^{(I)})$ is the Type IIA vector potential
on some Calabi-Yau manifold, then $g(L_{++}^{(I)})= 
f_V (L_{++}^{(I)})$ is the Type IIB hypermultiplet potential on the
same Calabi-Yau manifold.\foot{This relation can be proven using
sigma model arguments\eff. However, it does not follow directly from
mirror symmetry as was mistakenly claimed in \eff.}  

As shown in \eff using two-dimensional sigma model arguments, 
the vector and perturbative hypermultiplet contribution
to \act scales like $c^{-2}$ as one scales the string coupling-constant 
$\l_s \to c \l_s$, implying that they only
contribute at tree-level.
However, the hypermultiplet contribution to the action could also
get non-perturbative contributions which depend on Ramond-Ramond
zero modes, e.g.
\eqn\vvv{\int d^4 x|_{\t_j^\a=\tb_j^\ad=0} \oint_0 d\zeta 
(D_-)^2 (\bar D^+)^2 
(e^{-\tilde L^{(1)}/\zeta}~ 
f_T(\tilde L^{(J)},\zeta))
+~c.c.~.}

\newsec{Structure of N=2 D=8 Superspace}

The structure of superfields and actions in N=2 D=8 superspace is
almost identical to the structure in N=2 D=4 superspace, at least
for on-shell linearized superfields. This will allow the 
methods of the previous section to be repeated in this section. 

\subsec{N=2 D=8 Superfields} 

The variables of N=2 D=8 superspace are [$x^\mu$, $\t^\a_j$, $\tb^{j \ad}$]
where $\mu=0$ to 7, $\a$ and $\ad$ = 1 to 8, and $j=1$ to 2 is
an internal $SU(2)_R$ index which is raised and lowered using the
anti-symmetric $\e^{jk}$ tensor. $\tb^{j \ad }$ is the complex conjugate
of $\t^\a_j$, and under $U(1)_R$ transformations,
$\t^\a_j$ carries $+1$ charge and $\tb^{j\ad}$ carries $-1$ charge.

Under supersymmetry transformations parameterized by $\xi^\a_j$ and
$\bar\xi^\ad_j$, 
$$\d\t^\a_j =\xi^\a_j,\quad 
\d\tb^{j\ad} =\bar\xi^{j\ad},\quad 
\d x^\mu = i\s^\mu_{\a\ad}(\xi^\a_j\tb^{j\ad }  
+\bar\xi^{j\ad} \t_j^{\a }) ,$$ 
where $\s_{\a\ad}^\mu$ is the standard SO(8) Pauli matrix,
and supersymmetric derivatives are defined as 
\eqn\ww{
D_\a^j = {\p \over {\p \t^\a_j}} + i \tb^{j\ad } \s^\mu_{\a\ad} \p_\mu,\quad
\bar D^\ad_j = 
{\p \over {\p \tb_\ad^j}} + i \t_{j\a} \bar\s_\mu^{\ad\a} \p^\mu.}

The massless sector of the Type II superstring compactified on
$T^2$ is an N=2 D=8 supergravity multiplet\salam whose 128 bosonic fields
include 7 scalars, 6 vectors, 3 anti-symmetric two-forms, one
anti-symmetric three-form, and a spin-two graviton. 
At linearized level, these fields can be combined on-shell into a chiral 
superfield $W$ and a linear superfield $L_{jklm}$ 
which is symmetric
in its SU(2) indices and satisfies the reality condition 
$L_{jklm} = (L^{jklm})^*$. 

In addition to the chiral and linear
constraints
\eqn\addd{\bar D^j_\ad W = D^j_\a \bar W=0,\quad 
 D^\a_{(j} L_{klmn)} = 
\bar D^\ad_{(j} L_{klmn)} = 0,}
these superfields are related to each other by the constraint
\eqn\relat{D_{j}^\a D_{k}^\b \s^{\mu\nu}_{\a\b} W =  
\bar D^{l\ad} \bar D^{m\bd} \bar\s^{\mu\nu}_{\ad\bd} L_{jklm},\quad
\bar D_{j}^\ad \bar D_{k}^\bd \bar\s^{\mu\nu}_{\ad\bd} \bar W =  
 D^{l\a} D^{m\b} \s^{\mu\nu}_{\a\b} L_{jklm}.}
Note that $L_{++++}$ 
is a chiral/anti-chiral 
field 
satisfying
$\bar D^-_\ad L_{++++}$
$=$
$ 
D_+^\a L_{++++}$
$ = 0$ and 
\eqn\twtw{(D_-\s^{\mu\nu}D_-) L_{++++} = 
(\bar D^-\s^{\mu\nu}\bar D^-)\bar W,\quad
(\bar D^+\bar\s^{\mu\nu}\bar D^+) L_{++++} = 
(D_+\s^{\mu\nu}D_+)W.}

The physical bosonic fields appear in $W$ and $L_{jklm}$ as 
\eqn\phsdf{W = w + (\t_{j} \s^{\mu\nu} \t_{k}) F^{jk}_{\mu\nu} +
(\t_{j} \s^{\mu\nu\rho\kappa} \t^{j}) F_{\mu\nu\rho\kappa} 
+
(\t_{j} \s^{\mu\nu} \t_{k}) 
(\t^{j} \s^{\rho\kappa} \t^{k})
R_{\mu\nu\rho\kappa}) + ... ,}
$$L_{jklm} = 
l_{jklm} + (\t_j \s_{\mu\nu}\t_k) F^{\mu\nu}_{lm} 
+ (\tb_j \bar\s_{\mu\nu}\tb_k) \bar F^{\mu\nu}_{lm} 
+ (\t_j\s_{\mu\nu\rho}\tb_k) H^{\mu\nu\rho}_{lm} 
$$
$$+(\t_{j} \s^{\mu\nu} \t_{k}) 
(\tb_{l} \bar\s^{\rho\kappa} \tb_{m}) R_{\mu\nu\rho\kappa}
+ ...  ~~. $$
Under $U(1)_R \times SU(2)_R$, $W$ transforms as $(+4,1)$
and $L_{jklm}$ transforms as $(0,5)$, so the scalars
$w$ and $\bar w$ transform
as $(+4,1)$ 
and $(-4,1)$, the scalars
$l_{jklm}$ transforms as $(0,5)$, the vector field-strengths
$F_{jk}^{\mu\nu}$ and 
$\bar F_{jk}^{\mu\nu}$ 
transform as $(+2,3)$ and $(-2,3)$, the tensor
field-strength $H_{jk}^{\mu\nu\rho}$ transforms
as $(0,3)$, the self-dual and anti-self-dual part of the four-form
field-strength 
$F^{\mu\nu\rho\kappa}$ transform as $(+1,0)$ and $(-1,0)$, and the
curvature tensor 
$R_{\mu\nu\rho\kappa}$ transforms as $(0,1)$. 

The easiest way to 
understand the constraints of \relat
is to use the fact that the Type II closed
superstring field should be the `product' 
of two open superstring fields.\sieg
The massless
sector of the open superstring on $T^2$ is an N=1 D=8 super-Maxwell 
multiplet whose on-shell fields 
are described by a chiral and anti-chiral superfield, $\Phi$ and $\bar\Phi$,
which satisfy the constraint 
\eqn\sats{(D \s^{\mu\nu} D)\Phi = (\bar D\bar\s^{\mu\nu}\bar D)\bar\Phi.}
(In light-cone gauge, \sats is the familiar self-duality constraint\ref\lc
{L. Brink, O. Lindgren and B.E.W. Nilsson, Nucl. Phys. B212 (1983) 401\semi
S. Mandelstam, Nucl. Phys. B213 (1983) 149.}
$D^a D^b \Phi = \epsilon^{abcd}
\bar D_c \bar D_d \bar\Phi$ where $\t^a$ is an SU(4) spinor.)

So it is natural to interpret the chiral/chiral superfield
$W$ as the product $\Phi_L \Phi_R$ 
and the chiral/anti-chiral 
superfield $L_{++++}$ as the product $\Phi_L\bar\Phi_R$,
where 
$\Phi_L$ and $\Phi_R$ are
`left-moving' and `right-moving' open string fields, 
and $(\t_j^\a,\tb^j_\ad)$ split into left-moving
$(\t_-^\a,\tb^-_\ad)$ and right-moving
$(\t_+^\a,\tb^+_\ad)$. 
The constraint of \sats, when applied independently on $\Phi_L$ and $\Phi_R$,  
implies the constraints of \twtw,  
\eqn\stre{(D_-\s^{\mu\nu}D_-) L_{++++} = 
(\bar D^-\s^{\mu\nu}\bar D^-)\bar W, \quad
(\bar D^+\bar\s^{\mu\nu}\bar D^+) L_{++++} = 
(D_+\s^{\mu\nu}D_+)W.}

\subsec{N=2 D=8 Superspace Actions}

Although the constraint of \relat
puts $W$ and $L_{jklm}$ on-shell, one can ask
what kinds of supersymmetric actions can be constructed out of the on-shell
N=2 D=8 supergravity fields.
This analysis 
will be useful for proving non-renormalization theorems for the on-shell
S-matrix. 

Eight-derivative actions can be written in 
manifestly
N=2 D=8 supersymmetric notation as
\eqn\acttwo{\int d^8 x|_{\t_j^\a=\tb_j^\ad=0} [(D_+)^8 (D_-)^8 f_V(W) ~+ ~
\oint_0 d\zeta 
(D_-)^8 (\bar D^+)^8 f_T(\tilde L,\zeta) ~+~c.c.]}
where $f_V$ and $f_T$ are arbitrary functions, 
$\oint_0 d\zeta$ is a contour integration around
$\zeta=0$, and
\eqn\sdf{\tilde
L =  L_{++++} +\zeta L_{+++-} + \zeta^2 L_{++--} +
\zeta^3 L_{+---} + \zeta^4 L_{----}.}
The action is supersymmetric for the same reason as \act. 
Note that integrating over all 32 $\t$'s (i.e. using 
$(D_+)^8 (D_-)^8 
(\bar D^+)^8 (\bar D^-)^8 $) would imply a minimum of sixteen derivatives
in the action, assuming the action is local. As in the N=2 D=4 case, non-local
actions such as those coming from holomorphic anomalies
will not be discussed. 

Although \acttwo is written in terms of linearized superfields, 
note that the U-duality group is
$SL(3)\times SL(2)/SO(3)\times SO(2)$, and the internal automophism
group of the supersymmetry algebra is $SO(3)_R\times SO(2)_R$. 
This suggests that the full non-linear formulation of N=2 D=8
supergravity should include compensating scalars which 
parameterize 
$SO(2)\times SO(3)$, just like the scalars in the vector and
tensor compensators of the N=2 D=4 superstring effective action\eff. 

\newsec{Non-Renormalization Theorems for $R^4$ Terms}

The first step in 
proving non-renormalization theorems is to use the fact that
Ramond-Ramond zero modes decouple from perturbative amplitudes.
As will be shown later, the only Ramond-Ramond zero modes appearing
in $W$ and $L_{jklm}$ are the lowest components of $L_{+++-}$ and 
$L_{---+}$. Therefore, the perturbative contribution to the action 
of \acttwo must be invariant under 
$\tilde L \to \tilde L +  C \zeta + \bar C \zeta^3$ where
$c$ is a complex constant. 

When $f_T(\tilde L,\zeta)$ = $\zeta^{-1} g(\tilde L)$ for an
arbitrary function $g$, this is satisfied since the pole at
$\zeta=0$ is cancelled by the variation of $\tilde L$. 
This type of term contributes 
$$\int d^8 x|_{\t_j^\a=\tb_j^\ad=0} 
\oint_0 d\zeta \zeta^{-1}
(D_-)^8 (\bar D^+)^8 g(\tilde L) ~+~c.c.$$ 
$$=\int d^8 x|_{\t_j^\a=\tb_j^\ad=0}
(D_-)^8 (\bar D^+)^8 g(L_{++++}) ~+~c.c.$$ 
to the action of \acttwo. 

However, unlike the N=2 D=4 case,  
there is another type of
perturbative contribution from $L_{jklm}$ which is 
given by $f_T(\tilde L, \zeta) =h\zeta^{-3}\tilde L^5$ where $h$
is a constant.
Under $\tilde L \to  
\tilde L +  C \zeta + \bar C \zeta^3$, the variation proportional to $\bar C$
cancels the pole when $\zeta=0$. Furthermore, the term proportional to
$C$ does not contribute since 
$$\int d^8 x|_{\t_j^\a=\tb_j^\ad=0} 
\oint_0 d\zeta \zeta^{-2}
(D_-)^8 (\bar D^+)^8 \tilde L^4 ~+~c.c.$$ 
$$= \int d^8 x|_{\t_j^\a=\tb_j^\ad=0} 
\oint_0 d\zeta \zeta^{-2}
(\zeta^{-1} D_+)^8 (\zeta^{-1}\bar D^-)^8 \tilde L^4 ~+~c.c.$$ 
$$= \int d^8 x|_{\t_j^\a=\tb_j^\ad=0} 
\oint_0 d\zeta \zeta^{-18}
(D_+)^8 (\bar D^-)^8 \tilde L^4 ~+~c.c. ,$$ 
which vanishes since $\tilde L^4$ has a maximum of 16 $\zeta$'s, so
there is no term proportional to $\zeta^{-1}$. It is easy to check
that $f_T =\zeta^{-3} \tilde L^5$ is the only non-trivial
term of this type (e.g. when
$f_T = \zeta^{-3} \tilde L^4$, the action vanishes identically). 
 
So at the perturbative level, the only possible terms in the action are
\eqn\pert{\int d^8 x|_{\t_j^\a=\tb_j^\ad=0} [(D_+)^8 (D_-)^8 f_V(W) ~+ }
$$(D_-)^8 (\bar D^+)^8 \left( g(L_{++++}) ~+~ 
5h (L_{++++}^4 L_{++--} + 2 L_{++++}^3 L_{+++-}^2)\right) ] +~c.c.~.$$
In components, it is easy to compute that this gives 
\eqn\compo{\int d^8 x [ 
R^4_{++} ({\p\over \p w})^4 f_V(w) ~+ }
$$ R^4_{+-} 
\left( ({\p\over \p l_{++++}})^4 g(l_{++++}) ~+~ 
120 ~h  ~l_{++--}\right) +~c.c.~] + ...$$
where 
\eqn\defs{
R_{\pm\pm}^4 = (t^8 \pm {i\over 2}\e^8)^{\mu_1 \nu_1 ...\mu_4 \nu_4}
(t^8 \pm {i\over 2}\e^8)^{\rho_1 \kappa_1 ...\rho_4 \kappa_4} 
~\Pi_{n=1}^4 R^{\mu_n\nu_n\rho_n\kappa_n},}
$$(t^8 + {i\over 2}\e^8)^{\mu_1 \nu_1 ...\mu_4 \nu_4}
=\e^{\a_1 \b_1 ... \a_4 \b_4}~\Pi_{n=1}^4 \s^{\mu_n \nu_n}_{\a_n\b_n},$$
$$(t^8 - {i\over 2}\e^8)^{\mu_1 \nu_1 ...\mu_4 \nu_4}
=\e^{\ad_1 \bd_1 ... \ad_4 \bd_4}~\Pi_{n=1}^4 
\bar\s^{\mu_n \nu_n}_{\ad_n\bd_n},$$
and $...$ contains no $R^4$ terms. Note that $(t^8 \pm {i\over 2}\e^8)$ is
defined here with a D=8 Minkowski signature. 

At the non-perturbative level, Ramond-Ramond
zero modes do not have to decouple so one can have a term like
\eqn\dont{\int d^8 x [ 
R^4_{+-}  e^{-l_{++--}}({\p\over{\p l_{++++}}})^4 f(l_{++++}) + ~c.c.~] }
which comes from the superspace expression 
\eqn\donts{\int d^8 x|_{\t_j^\a=\tb_j^\ad=0} 
\oint_0 d\zeta \zeta^{-1} (D_-)^8 (\bar D^+)^8 \left( f(\tilde L) 
e^{-\tilde L/\zeta^2} \right) + ~c.c.~ .}

To determine the loop-order of the terms in \pert and to determine which
terms survive in the uncompactified limit, one needs to know how
the scalar fields depend on the string coupling constant and on the
$T^2$ volume. Although the scalars $l_{jklm}$ and $w$ are
only defined at linearized level, this will be enough to prove
non-renormalization theorems for the $R^4$ terms. 

The seven scalar moduli consist of the complex modulus $U= U_1 +i U_2$ of
the two-torus, the kahler modulus $T = T_1 +i T_2$ of the
two-torus ($T_2$ is the volume), the complex 
Ramond-Ramond scalar $B= B_1 + i B_2$, 
and the D=8 string coupling constant
$\l_8$ which is related to the D=10 string
coupling constant by $\l_{8}=
(T_2)^{-\half} \l_{10}$. 

These moduli for the Type IIB superstring
can be combined into the following symmetric
matrices with determinant one\ref\kir
{E. Kiritsis and B. Pioline, ``On $R^4$ Threshold Corrections
in IIB String Theory and (p,q) String Instantons'', hep-th/9707018.}
\eqn\ms{M_1 = {1\over U_2}\left(\matrix{1&U_1\cr
U_1&|U|^2\cr}\right) , ~
M_2 = {1\over{(\l_8)^{2/3} T_2}}\left(\matrix{1 &T_1
&-B_1\cr T_1
&|T|^2
&{\rm Re}(\bar T B) \cr  -B_1
&{\rm Re}(\bar T B)&
{{T_2}(\l_8)^2}+|B|^2\cr }\right)}
which transform as $M_a \to \Omega_a M_a \Omega_a^T$
under the SL(2,R) and SL(3,R) transformations generated by
$\Omega_1$ and $\Omega_2$. 
For the Type IIA superstring, the only difference is
that the $T$ and $U$ moduli switch places. 

Expanding to first order near $M_1 =M_2 = 1$ (i.e. $T=U=i$,
$\l_8=1$, $B=0$), one finds 
\eqn\mstwo{M_1 = \left(\matrix{1-\hat U_2 &\hat U_1\cr
\hat U_1&1 +\hat U_2\cr}\right) , \quad 
M_2 = \left(\matrix{1-\hat T_2 -{2\over 3}\hat \l_8 &\hat T_1
&-\hat B_1 \cr \hat T_1
& 1+\hat T_2 -{2\over 3}\hat\l_8 
& \hat B_2 \cr  -\hat B_1 
&\hat B_2 &
1+{4\over 3} \hat\l_8\cr
}\right) }
where $\hat U=U-i$, $\hat T= T-i$, $\hat \l_8= \l_8 -1$, and $\hat B=B$. 
Using the transformation properties of these
matrices under $SO(2)\times SO(3)$ and comparing with
the $U(1)_R\times SU(2)_R$ charges of $w$ and $l_{jklm}$, one
learns for the Type IIB superstring that 
\eqn\rela{w= \hat U ,\quad l_{++++} = \hat T ,\quad
l_{+++-} = \hat B,\quad l_{++--} = \hat \l_8 }
where the $SU(2)_R$ transformations are defined by commuting $M_2$ 
with 
$$J_x =\left(\matrix{0 &0 
&0 \cr 0
& 0
& 1 \cr  0 
& -1 &
0\cr
}\right), ~
J_y =\left(\matrix{0 &0 
&1 \cr 0
& 0
& 0 \cr  -1
& 0 &
0\cr
}\right), ~
J_z = \left(\matrix{0 & 1
&0 \cr -1
& 0
& 0 \cr  0 
&0 &
0\cr
}\right).$$
Note that in the Type IIB (Type IIA) superstring, 
the lowest component of the chiral/chiral
spacetime superfield $W$ describes the chiral/chiral (chiral/anti-chiral) 
worldsheet modulus of $T^2$, while the 
lowest component in the chiral/anti-chiral
superfield $L_{++++}$ 
describes the chiral/anti-chiral (chiral/chiral) worldsheet
modulus of $T^2$. 

So near $\hat U=\hat T=\hat B=\hat \l_8=0$, the $R^4$ terms in the
effective action of 
the Type IIB superstring appear as 
\eqn\near{\int d^8 x [ 
R^4_{++} ({\p\over \p \hat U})^4 f_V(\hat U) ~+ }
$$ R^4_{+-} 
\left( ({\p\over \p \hat T})^4 g(\hat T) + 
24 h  \hat \l_8 +  C(\hat T,\hat \l_8)
\right) +~c.c.~]$$
where $C(\hat T,\hat \l_8)$ comes from non-perturbative contributions. 
The terms proportional to $f_V$ and $g$
come from one-loop since they are independent
of $\l_8$, while the loop dependence of the term proportional to $h$
can not
be determined from a linearized analysis. However, since there
is precisely one $\l_8$-dependent perturbative term at linearized level, 
there should be precisely one $\l_8$-dependent term in the
full non-linear action. 

Therefore, in the full non-linear effective
action of the Type IIB superstring on
$T^2$, the $R^4$ terms appear as   
\eqn\full{\int d^8 x \sqrt{g_8} [ 
R^4_{++}  A(U) + 
R^4_{+-} 
\left( B(T) + 
h_g  (\l_8)^{2g-2} +  C(T,\l_8)
\right) +~c.c.~]}
where $h_g$ is a constant. For the Type IIA superstring,
the $T$ and $U $ moduli exchange places. 

Up to terms coming from the holomorphic anomaly, this is in
precise agreement with the results of \kir where 
$$A(U)=4\pi\log \eta(U),\quad B(T)=4\pi\log\eta(T),\quad h_0 =2\zeta(3).$$ 
Note that $A$ and $B$ are the same function, which does not
follow from $T$-duality. It is analogous to the Type IIA/Type IIB
relation between $f_T$ and $g$ in the N=2 D=4 action which was
mentioned in footnote 1.

To determine $R^4$ terms in the effective action of the uncompactified
Type II superstring, one needs to take $T_2$ to infinity and keep
terms which diverge linearly with $T_2$, remembering that $\l_8$ scales
like $T_2^{-\half}$. For the Type IIB superstring, the only terms
which survive from \full are  
\eqn\surv{\int d^{10} x \sqrt{g_{10}} 
R^4_{+-} [h_0  (\l_{10})^{-2} +  
\lim_{T_2\to\infty} T_2^{-1} ( B(T) + 
C(T,\l_{10} T_2^{-\half})] + ~c.c.~,}
so $R^4$ terms only get tree-level, one-loop and non-perturbative 
contributions. For the Type IIA superstring, the only terms which survive are
$$\int d^{10} x \sqrt{g_{10}} \left(  
R^4_{+-} h_0  (\l_{10})^{-2} +  
R^4_{++}\lim_{T_2\to\infty} T_2^{-1} A(T) \right) + ~c.c.~,$$ 
so $R^4$ terms only get tree-level and one-loop contributions. 

\newsec{Possible Generalizations of the Non-Renormalization Theorems}

In this paper, $R^4$ terms in the effective action of the 
Type II superstring compactified on $T^2$ were constructed in
N=2 D=8 superspace, which allowed a simple proof of $R^4$
non-renormalization theorems. This superspace
construction was very similar
to the construction of the vector and hypermultiplet potentials
in the N=2 D=4
superspace effective action of the D=4 Type II superstring. 

As is well-known, there are higher-derivative topological amplitudes\hol
of the D=4 Type II superstring which have properties similar
to those of the vector and hypermultiplet potentials. These
topological amplitudes come from terms which can be written in
N=2 D=4 superspace as 
\eqn\zzz{\int d^4 x|_{\t_j^\a=\tb_j^\ad=0} [(D_+)^2 (D_-)^2 
\left( (P_{\a\b} P^{\a\b})^{g} A_g(W^{(I)})\right) +}
$$ 
(D_-)^2 (\bar D^+)^2
\left((Q_{\a\bd} Q^{\a\bd})^{g} 
B_g(L_{++}^{(J)})\right) ~+~c.c.]$$ 
where $P_{\a\b}$ and $Q_{\a\bd}$ are chiral and twisted-chiral
field-strengths constructed from the supergravity multiplet.
In components, these terms are 
$$\int d^4 x \sqrt{g_4}[R^2 (F_{\mu\nu} F^{\mu\nu})^{g-1} A_g(w^{(I)}) +  
R^2 (\p_\mu Z \p^\mu Z)^{g-1} B_g(l_{++}^{(J)}) ~+~c.c.] + ...$$ 
where $F^{\mu\nu}$ is the graviphoton field-strength and $Z$ is
a complex Ramond-Ramond scalar. Using arguments similar to those
of section 2, one can prove 
that $A_g$ only appears at genus $g$ and $B_g$
only appears at genus $g$ and non-perturbatively.\eff  

It is less well-known that there are also higher-derivative
topological amplitudes of the D=8 Type II superstring which have
properties similar to those of the $R^4$ term. These amplitudes
were first shown to be topological for the Type II superstring
compactified on $K3$\ref\topo{N. Berkovits
and C. Vafa, Nucl. Phys. B433 (1995) 123, hep-th/9407190.}, 
and were later computed explicitly\ref\oog
{H. Ooguri and C. Vafa, Nucl. Phys. B451 (1995) 121, hep-th/9505183.}
for the Type II superstring compactified on $T^2\times R^2$
(i.e. for the Type II superstring compactified on $T^2$). 

For the Type IIA superstring, topological amplitudes at genus $g$
come from terms of the form
\eqn\eigh{\int d^8 x \sqrt{g_8}[R^4 (F^4)^{g-1} A_g(T,\bar T) +  
R^4 (H^4)^{g-1} B_g(U,\bar U) ~+~c.c.] + ... }
where $F^4$ is constructed from the vector field-strengths and
$H^4$ is constructed from the tensor field-strengths. In the language
of \oog, this amplitude comes from the top instanton number and
\eqn\newa{A_g(T,\bar T) = (\l_8)^{2g-2} (T_2)^g \sum_{(m,n)\neq(0,0)}
{{|n+m T|^{2g-4}}\over (n+m T)^{4g-4}},}
$$
B_g(U,\bar U) = (\l_8)^{2g-2} (U_2)^g \sum_{(m,n)\neq(0,0)}
{{|n+m U|^{2g-4}}\over (n+m U)^{4g-4}}.$$
For the Type IIB superstring, $T$ and $U$ exchange places.

As was first pointed out by Vafa\ref\private{C. Vafa, private
communication.}, this suggests there might
be non-renormalization theorems for $R^4 F^{4g-4}$ and
$R^4 H^{4g-4}$ terms in the Type II superstring effective action. 
Note that these terms come from dimensional reduction of 
$R^{2g+2}$ terms in eleven dimensions. At this moment,
it is not known how to write the action of \eigh in N=2 D=8 superspace,
so the methods of this paper can not be used to check the existence
of these non-renormalization theorems. 

One mysterious feature of such a theorem is that it would naively imply
that $R^{2g+2}$, like $R^4$, recieves contributions only at genus
$g$ and below in the effective action of the 
uncompactified Type IIA superstring. Note that $A_g$ scales like
$T_2$ as $T_2\to\infty$, so this term appears to be 
present in ten dimensions, and to blow
up like $(R_{11})^{g-1}$ in eleven dimensions. 
As pointed out in
\ref\tseytlin{J.G. Russo and
A.A. Tseytlin, ``One-Loop Four Graviton Amplitude
in Eleven-Dimensional Supergravity'', hep-th/9707134.}, 
lack of non-perturbative corrections to $R^{2g+2}$ terms
would violate eleven-dimensional covariance of $M$-theory 
since only $R^{3g+1}$ terms can come
from dimensional reduction of Lorentz-covariant terms in eleven dimensions.

Note
that if $R^{2g+2}$ were a topological term in the effective
action of the uncompactified Type II superstring
action, it would be reasonable for $F^{2g+2}$ to be 
a topological term in the effective action of the Type I superstring, since
the photon vertex operator is the `square-root' of the graviton
vertex operator. The topological nature of such an $F^{2g+2}$ term
is supported by recent M(atrix) model computations\ref\becker{
K. Becker, M. Becker, J. Polchinski and A. Tseytlin, 
Phys. Rev. D56 (1997) 3174,
hep-th/9706072.}
\ref\md{M. Douglas and H. Ooguri, private communication.}.
\vskip 20pt

{\bf Acknowledgements:} I would like to thank Cumrun Vafa for
suggesting that the N=2 D=8 effective action resembles
the N=2 D=4 effective action and that similar
methods might be used to prove non-renormalization theorems.
I would also like to thank B. de Wit, M. Douglas, M. Green, P. Howe,  
H. Ooguri, W. Siegel, and
P. Townsend 
for useful conversations.This
work was partially supported by 
CNPq grant number 300256/94-9.

\listrefs
\end